\documentclass[10pt]{article}
\usepackage[dvips]{graphicx,graphics}
\usepackage{mmr}
\newcommand{\falpha}{\mbox{\boldmath$\alpha$}}
\newcommand{\fbeta}{\mbox{\boldmath$\beta$}}
\newcommand{\fomega}{\mbox{\boldmath$\omega$}}
\newcommand{\bx}{{\bf x}}
\newcommand{\bpi}{{\bf p}}
\newcommand{\bPi}{{\bf\Pi}}
\newcommand{\bC}{{\bf C}}

\title{Resampling Approach to the Estimation of \\Reliability Systems}
\author{\bf {Maxim Fioshin, Helen Fioshina}
\\
Institute of Transport Machines and Technologies,\\
Riga Technical University,\\
Lomonosova Str. 1, LV1019, Riga\\
Latvia\\
{\em Maksims.Fiosins@rtu.lv, Jelena.Fiosina@rtu.lv } 
}

\begin{document}
\maketitle 
\begin{abstract}
The article is devoted to the resampling approach application to 
the reliability problems. This approach  to
reliability problems was first proposed by Ivnitsky (1967). Resampling is 
intensive statistical computer method, which is non-parametrical, that uses initial
samples data in different combinations to simulate the process many times and get finally the estimator of the characteristics of interest.
At the present paper simple resampling, hierarchical resampling, the case of one sample for several variables, the case of partially known distributions, analysis of degradation flow, analysis of degradation-renewal process, construction of confidence intervals are described. All those resampling application cases can be applied successfully to solve the reliability problems as an alternative to classical methods.\footnote{This work has been partly supported by the European Social Fund within the 
National Programme "Support for the carrying out doctoral study program's
and post-doctoral researches" project "Support for the development of doctoral
studies at Riga Technical University"}
\end{abstract}

\section*{Introduction}
Computers play an important role in the development of modern statistical science. Many statistical methods developed in the last time require a big amount of computations. Such methods are usually called intensive statistical methods. Some authors speak about separate discipline called "computational statistics" \cite{Gentle_02}. 

In statistical problems of reliability we often have difficulties with application of classical methods. In the case of complex systems, small samples, unknown distributions of system characteristics classical methods work not so good. In this case, methods of computational statistics can be used as an alternative.

The field of computational statistics at present time includes a big number of methods. 
Jackknife method was suggested by M. Quenouille in 1949 \cite{Quen_49}. It uses the estimator, which is the combination of the estimator's obtained using all data and the estimators, obtained using only part of the same data. In 1967 V. Ivnitsky \cite{Ivn_67} suggested to use resampling for the estimation of the systems' reliability by simulation. B. Efron \cite{Efron_79_2} suggested bootstrap method, which is the generalization of the jackknife method. Resampling approach was investigated from 1995 under prof. A. Andronov supervision. During this investigation, simple and hierarchical resampling \cite{Andr_95} and their implementations in reliability theory, queuing theory \cite{AndrF_99_2,Afan_02}, stochastic processes \cite{Andr_00_2,Afan_05_1,Andr_06}, optimization tasks \cite{Andr_00} and construction of confidence intervals \cite{Andr_02,AndrF_04} were considered. The present paper is devoted to the description of the main
results connected with the reliability problems.

The classical "plug-in" approach consists in the choosing of the forms of unknown distributions and estimating their unknown parameters on the base of available sample populations. Then the estimated distributions are used in formulas for calculation of unknown parameters instead of real values or they are used for generation of the pseudo-random numbers, which are used in simulation instead of latter variables. Here the estimation of probability distributions, espetially in the case of small initial samples, leads to making mistakes in choosing  the form of a distribution and estimating its parameters. It can lead to the bias of estimated parameters and results of the simulation.

Resampling is an alternative approach. We do not estimate unknown distributions, but extract values from initial sample populations at random and use them as values of input variables during simulation. This method does not require any preliminary information about the type of the distributions, so it is non-parametrical. It uses the available information in different combinations, which require a big number of computations, but allows to get good estimators in situations, where classical estimators are not good.

The first section of the present article is devoted to the main principles of the resampling approach. The resampling algorithm and the principals of the resampling estimator calculation are described. The variance of the estimator is supposed to be criterion of obtained estimator efficiency. In the case of biased estimators it is better to take the mean squared error as the criterion of efficiency, which is calculated on the base of variance and bias. The variance is calculated on the base of resampling pairs ($\fomega$-pair, $\fbeta$-pair, $\falpha$-pair).
The resampling pair shows the common elements structure of two different resamples, so the type of the pair depends on the problem. In this section the resampling is applied to the estimation of the reliability function of logical schema. Different variants of this method are described: simple resampling, hierarchical resampling, resampling of one sample for several distributions and the case of partially known distributions.

In the second part the resampling approach implementation to stochastic processes analysis is described. Two problems are  considered there. The first problem is analysis of the degradation process with the accumulation of damages. The second problem is the comparison of two renewal processes implementing to the analysis of degradation-renewal process analysis.  

The last part of the article is devoted to resampling application to the interval estimation of logical system and the algorithm of true coverage probability calculation of the constructed confidence interval.

Finally the concluding remarks are made. All the parts contain only the brief description of the considered problems and the main results. For more detailed description please refer to a bibliography, which is presented at the end of the article.

\section{Resampling estimation of reliability function of logical schemes}

\subsection{Simple resampling}

Let we have a known function $\phi(x_1,x_2,\ldots,x_m)$ of $m$ real arguments. This function can represent some characteristics
of $m$-element logical system with working times of elements $x_1,x_2,\ldots,x_m$ (for example, the total working time of the system, the indicator that the system works at the given time $t$ etc.). 

Let we have a vector of $m$ independent r.v. {\bf X}=($X_1$,$X_2$,\ldots,$X_m$), each component with unknown cdf $F_i(x)$; only sample populations {\bf H}=($H_1$,$H_2$,\ldots,$H_m$), $H_i=(X_{i,1},X_{i,2},\ldots,X_{i,n_i})$ are available for each $X_i$, $n_i=|H_i|$. 
 Let our parameter of interest is the expectation of the function $\phi$, which argument is vector {\bf X}: 
\begin{equation}
\label{eq_2_1}
\Theta = E\;\phi({\bf X}).
\end{equation}
We have to estimate $\Theta$ on the base of given sample populations {\bf H}.

The classical nonparametric {\it plug-in} approach is the following. We calculate the empirical cdf $\hat{F}_i(x)$, $i$=1,2,\ldots,$m$ on the base of samples {\bf H}, take formulas for calculation of $\Theta$ (which include $F_i(x)$) and use $\hat{F}_i(x)$ instead of $F_i(x)$. This gives formula for {\it plug-in} estimator $\hat{\Theta}$ of $\Theta$. Usually this estimator has good properties, but in some cases (the same samples for the same variables, small samples, etc.) it can have big variance or be biased.

The {\it resampling} approach is nonparametric approach to systems simulation and estimation \cite{Andr_00}.
It is a simple alternative to classical methods. We perform the following iterative procedure. On the $q$-th step we extract at random one element $X^{*q}_i$ from each population $H_i$, $i=1.2\ldots,m$. 
Let ${\bf j}(q)=\{j_1(q),j_2(q),\ldots,j_m(q)\}$ be the indices of extracted elements in the corresponding populations {\bf H}, so $X^{*q}_{i}=X_{i,j_i(q)}$.
The estimator $\Theta^{*q}$ of $\Theta$ is calculated using vector ${\bf X}^{*q}=\{X^{*q}_1,X^{*q}_2,\ldots X^{*q}_m\}$ as argument of the function $\phi$:
\begin{equation}
\label{eq_2_2}
\Theta^{*q} = E\;\phi({\bf X}^{*q}).
\end{equation}

The procedure is repeated $r$ times, and as the estimator $\Theta^*$ of $\Theta$ an average of all $\Theta^{*q}$ is taken: 
\begin{equation}
\label{thetaest}
\Theta^* = \frac{1}{r} \sum_{q=1}^r \Theta^{*q}.
\end{equation}

The obtained estimator (\ref{thetaest}) is unbiased:
\begin{equation}
\label{unbiased}
E\;\Theta^* = \Theta.
\end{equation}

So, in order to provide proper application of {\it resampling} method, we should know other properties of $\Theta^*$. In most cases estimators' variance $Var\; \Theta^*$ can be taken as efficiency criterion. Although, in some situations the estimator $\Theta^*$ is biased and then bias and mean squared error $MSE\;\Theta^*$ also should be calculated.

The variance $Var\;\Theta^*$ of the {\it resampling} estimator $\Theta^*$ is:
\begin{equation}
\label{thetavar}
Var\;\Theta^* = \frac{1}{r}\;Var\;\Theta^{*q} + \frac{r-1}{r} \; Cov\;(\Theta^{*q},\Theta^{*q'})=
\frac{1}{r}\;\mu_2+\frac{r-1}{r}\;\mu_{11}-\mu^2, \qquad q\ne q',
\end{equation}
where $\mu_2=E\;([\Theta^{*q}]^2)$, $\mu_{11}=E\;\Theta^{*q}\Theta^{*q'}$ $(q\ne q')$, $\mu=E\;\Theta^*$.

In this formula, the variance $Var\;\Theta^{*q}$, the second moment $\mu_2$ and the expectation
$\mu$ depend only on the properties of the function $\phi$ and r.v. {\bf X}.
Only the covariance $Cov(\Theta^{*q},\Theta^{*q'})$ and the mixed moment $\mu_{11}$ depend on the applied {\it resampling} procedure. So our goal is the calculation of the variance (\ref{thetavar}) depending on the used {\it resampling} procedure.

In order to calculate the covariance $Cov(\Theta^{*q},\Theta^{*q'})$ or the mixed moment $\mu_{11}$, we introduce
the notation of the $\fomega$-pair.  

Let ${\bf M}=\{1,2,\ldots,m\}$. We say that two vectors
${\bf j}(q)$ and ${\bf j}(q')$ produce the $\fomega$-pair, $\fomega\subset M$, when
$j_i(q)=j_i(q')$, if $i\in\fomega$ and $j_i(q)\ne j_i(q')$, if $i\notin\fomega$. So, the
$\fomega$-pair shows indices of the arguments which have the same elements from the initial
sample populations in two different realizations $q$ and $q'$.

For example, let we have function $\phi(x_1,x_2,x_3,x_4,x_5)$ and the resample vectors {\bf j}$(q)=(4,1,2,3,3)$ and {\bf j}$(q')=(1,2,2,4,3)$. In this case, they will form $\fomega=(3,5)$-pair.

Let us denote $A_{\fomega}$ the event "$\fomega$-pair is happened". Let $P\{\fomega\}$=$P\{A_{\fomega}\}$ be the probability of this event.
We can calculate the covariance $Cov(\Theta^{*q},\Theta^{*q'})$ or the mixed $\mu_{11}$ given 
$A_{\fomega}$.

Let $C({\fomega})=Cov(\Theta^{*q},\Theta^{*q'})|A_{\fomega}$, $q\ne q'$ be conditional covariance 
given {$\fomega$}-pair.
Let $\mu_{11}({\fomega})=\mu_{11}|A_{\fomega}$, $q\ne q'$ be conditional mixed moment given
{$\fomega$}-pair.
Then $Cov(\Theta^{*q},\Theta^{*q'})$ can be calculated as follows:
\begin{equation}
\label{cov}
Cov(\Theta^{*q},\Theta^{*q'})=\sum_{{\fomega}\in M} P\{{\fomega}\}C({\fomega}).
\end{equation}

The mixed moment $\mu_{11}$ can be calculated as follows:
\begin{equation}
\label{mu11}
\mu_{11}=\sum_{{\fomega}\in {M}} P\{{\fomega}\}\mu_{11}({\fomega}).
\end{equation}

Now let us show how to obtain the probabilities $P\{\fomega\}$.
As the arguments of the function $\phi$ are independent and 
the probability to extract the same element from $H_i$ twice on the $q$-th realization 
and on the $q'$-th realization is $1/n_i$, the probability $P\{\fomega\}$ can
be calculated as follows:
\begin{equation}
\label{omegapr}
P\{\fomega\}=\prod_{i\in\fomega}\frac{1}{n_i} \prod_{i\notin\fomega}\left(1-\frac{1}{n_i}\right).
\end{equation}

The construction of all $\fomega$-pairs is simple combinatorial problem.
By constructing all the $\fomega$-pairs we can calculate the variance (\ref{thetavar}), using formulas ($\ref{cov}$)-($\ref{mu11}$).

As an example, let us consider a system "2 of 3" \cite{Fio_03}.
It consists on 3 elements, and it works if at least 2 of 3 elements work. 
The problem is to estimate the probability that at the time moment
$t$ the system works.

The function of interest is the following indicator function $\phi_t(x_1,x_2,x_3)$:
$$
\phi_t(x_1,x_2,x_3)=\left\{\begin{array}{rl}
1 &\mbox{if at least 2 elements of $\{x_1,x_2,x_3\}$ are greater than $t$},\\
0 &\mbox{otherwise. }
\end{array}\right.
$$

Our aim is to estimate the probability that system works at time moment $t$ as the expectation of the function: $
\Theta_t = E\;\phi_t(x_1,x_2,x_3).
$

In our case the following 8 $\fomega$-pairs are possible: $
\{\},\{1\},\{2\},\{3\},\{1,2\},\{1,3\},\{2,3\},\{1,2,3\}
$

The probabilities of the $\fomega$-pairs and the mixed moments $\mu_{11}(\fomega)$ can be easily calculated. This allows us to use the resampling approach and to calculate 
the variance of obtained estimator.

\subsection{Hierarchical resampling}

Often the simulated system is complex, but structured. It can be split onto subsystems, which can be simulated separately. Then, the results of subsystems simulation 
can be used to simulate the whole system.
In this case we can use {\it hierarchical} resampling. 

{\it Simple resampling} can also be used
for analysis of hierarchical systems. We can extract, like in previous section,
values for input data from initial samples, calculate the value of function of
interest without paying attention to its hierarchical structure and get the
simple resampling estimator. But {\it hierarchical} resampling has some advantages
in the described situation \cite{Andr_95} and allows: parallel calculations, sample sizes optimization, more clear analysis of the efficiency. 

Let the function $\phi$ has hierarchical structure,
i.e. it can be split into subfunctions $\phi_i$. The results of subfunctions are 
used as arguments of the functions on the higher layer. 

It is convenient to represent function $\phi$ by the calculation tree. 
The root of this tree has index $k$ and corresponds to the function
$\phi=\phi_k$. Nodes with indices $1, 2, \ldots, m$
correspond to input variables $X_1, X_2, \ldots, X_m$. The rest nodes
are intermediate ones, which correspond to intermediate functions
$\phi_{m+1}, \phi_{m+2}, \ldots, \phi_{k-1}$.

Let us denote $I^v$ the set of nodes, from which 
arcs go to the node $v$, 
and $I_0^v$ - the set of initial variables, such that
the node $v$ depends upon them. Note that for vertices of the same level 
the sets $I^v$ do not cross and the sets $I_0^v$ also do not cross.

Now, the function $\phi$ can be calculated by so-called {\it "wave"} algorithm \cite{Andr_95}. 
We create samples $H_{m+1}, H_{m+2}, \ldots, H_k$ sequentially. The sample $H_v=(X_{v,1},X_{v,2},\ldots,X_{v,n_v})$, $v=m+1,m+2,\ldots,k-1$ 
is calculated by extracting values from corresponding samples $H_i, i\in I_v$ and using them as arguments of the function $\phi_v$. It is clear that
\begin{equation}
\label{hierelem}
X_{v,q}=\phi(X_i^{*q}|i\in I^v), q=1,2,\ldots,n_v.
\end{equation}

Finally, the estimator $\Theta^*$ of $\Theta$ can be calculated as an average of the sample $H_k$ elements at the root 
of the calculation tree:
\begin{equation}
\Theta^{*}=\frac{1}{n_k}\sum_{i=1}^{n_k} X_{k,i}.
\end{equation}

Now our purpose is to calculate variance $Var\;\Theta^*$. It can be calculated by formulas (\ref{thetavar})-(\ref{mu11}), but 
using slightly generalized $\fomega$-pair notation.

Let us consider the value $X_{v,q}$, calculated by formula (\ref{hierelem}). 
It was calculated using {\it only one} value from each sample $H_i$, $i\in I_0^v$.
Let us define these values indices in the initial samples by vector ${\bf j}^v(q)=\{j_i^v(q)\}$, $i\in I_0^v$. 
Some elements of vectors ${\bf j}^v(q)$ and ${\bf j}^v(q')$ can be equal. So, we can use almost the same
definition of the $\fomega$-pair.

Let {\bf M}=\{1,2,\ldots,$m$\}. 
We will say that two vectors
${\bf j}^v(q)$ and ${\bf j}^v(q')$ produce the $\fomega$-pair, $\fomega\subset M$, when
$j_i^v(q)=j_i^v(q')$, if $i\in\fomega$ and $j_i^v(q)\ne j_i^v(q')$, if $i\notin\fomega$.

Let $A^v_{\fomega}$ be the event "the $\fomega$-pair takes place at the node $v$". Let $P^v\{\fomega\}$ be the probability of this event. The values of $P^v\{\fomega\}$ is calculated recurrently. If the node
$v$ is on zero level $(1\le v\le m)$, then $P^v\{{\fomega}\}$ can
be calculated easily:
\begin{equation}
P^v\{\fomega\}=\left\{
\begin{array}{ll}
0&\mbox{if }\;\fomega\ne\emptyset,\\
1&\mbox{otherwise.}
\end{array}
\right.
\end{equation}

Now let us consider a node $v$ on another level. Let us consider
a sample $H_i$, $i\in I^v$ from which the node $v$ depends.
The probability that some element from $H_i$ is
chosen twice: for $X_{v,q}$ and for $X_{v,q'}$ is equal to $1/n_i$.

The event $A^v_{\fomega}$ can happen if in each node $i\in I^v$
one of two events occurs:
\begin{itemize}
\item We selected different elements from $H_i$ for $X_{v,q}$ and $X_{v,q'}$
(with the probability $1-1/n_i$) 
and selected elements produced $\fomega\cap I_0^i$ - pair (with the probability $P^i\{\fomega\}$);
\item We selected the same element from $H_i$ for $X_{v,q}$ and $X_{v,q'}$
(with the probability $1/n_i$). In this case all elements from $I_0^i$ will
be extracted twice. The event $A^v_{\fomega}$ can happen in this case only
if all elements of $I_0^i$ belong to $\fomega$: $I_0^i\subset\fomega$.
\end{itemize}

Now let us denote $\delta_{i,\fomega}$:
\begin{equation}
\delta_{i,\fomega}=\left\{
\begin{array}{ll}
1&\mbox{if}\;I_0^i\subset\fomega,\\
0&\mbox{otherwise.}
\end{array}
\right.
\end{equation}

Now we can write formula for the calculation of $P^v\{\fomega\}$:
\begin{equation}
P^v\{\fomega\} = \prod_{i\in I^v} \left((1-1/n_i)P^i\{\fomega\}+(1/n_i)\delta_{i,\fomega}\right).
\end{equation}

Our goal is to calculate $P^k\{\fomega\}$ for all $\fomega\in {\bf M}$. This
will allow us to use formula (\ref{cov}) or (\ref{mu11}) for calculation
the value $\mu_{11}$ or $Cov(\Theta^{*q},\Theta^{*q'})$. Then we will use (\ref{thetavar}) for calculation
of the estimator $\Theta^*$ variance.

Let us consider an example. Let we have a system that consists of 6 elements. 
The 1-st and the 2-nd elements are connected in parallel, the 3-d and the 4-th elements are connected 
sequentially, the 5-th and the 6-th elements are connected in parallel, but the 6-th element is switched on when the 5-th element
fails (cold reservation). Our purpose is to estimate the probability that the system will work at the time moment $t$.

Our function of interest $\phi_t$ can be denoted as follows:
\begin{equation}
\phi_t(x_1,x_2,x_3,x_4,x_5,x_6)=
\left\{
\begin{array}{ll}
1,& \mbox{if }min(max(x_1,x_2),min(x_3,x_4),x_5+x_6)<t,\\
0,& \mbox{otherwise}.
\end{array}
\right.
\end{equation}

The function $\phi_t$ can be represented
by using the calculation tree. The tree has 6 leaves, three elements on the first level correspond to the
functions $max(x_1,x_2)$, $min(x_3,x_4)$ and $x_5+x_6$ and the element on the second level (root of this tree)
corresponds to the function $min$. 

In this case we have 4 $\fomega$-pairs in each node on the first level and 8 $\fomega$-pairs in the root of the tree. This allows us using the resampling approach and calculating
the variance of obtained estimator. 



\subsection{The case of the only one sample for several r.v.}

In this section we will show how to deal with in the case, when only one  sample is available for several logical elements.
This situation can often appear when elements are considered to have identical characteristics.

Let we have a function $\phi(x_1,x_2,\ldots,x_m)$, but some arguments are considered to be statistically identical.
We have only one sample for all identical arguments. Let the numeration of arguments corresponds to the order of 
sample numbers. So, let we have samples $H_1,H_2,\ldots,H_k$ and the sample $H_i$ is used for arguments $(x_{l_{i-1}+1} \ldots x_{l_i})$ (and we put $l_0=0$). Let $m_i$ be a number of arguments, for which the sample $H_i$ is used, $m_i = l_i-l_{i-1}+1$. 

The {\it resampling} approach here is the following: on the $q$-th realization we extract one value for each argument from the corresponding population. The values from one sample are extracted without replacement. The extracted values are used as arguments of the function $\phi$, and the resampling estimator for $q$-th realization $\Theta^{*q}$ is calculated by formula (\ref{eq_2_2}). Finally, the resampling estimation $\Theta^{*}$ is calculated by formula (\ref{thetaest}). Note that obtained estimator is unbiased. Our goal is to show how to calculate its variance.

The variance $Var\;\Theta^{*}$ is calculated by formula (\ref{thetavar}). In order to calculate the covariance 
$Cov(\Theta^{*q},\Theta^{*q'})$ we generalize the notation of $\fomega$-pair, introducing the $\fbeta$-pair. The idea lying behind
this definition is the following: in the previous cases, where $\fomega$-pair was used, the same element from initial sample in two different realizations $q$ and $q'$ was
possible only for the argument $x_i$. In the present case the same element in two different realizations $q$ and $q'$ is possible for different arguments, namely any of arguments $(x_{l_{i-1}+1} \ldots x_{l_i})$ can have the same element, because it is extracted from $H_i$ for all of them. The $\fbeta$-pair shows what arguments have the same elements extracted.

Let ${\bf j}(q) = (j_1(q),j_2(q),\ldots,j_m(q))$ be the indices of elements extracted on the $q$-th step. Note that $(j_{l_{i-1}+1}(q) \ldots j_{l_i}(q))$ are indices in the sample $H_i$ and they are different, $i=1,\ldots,k$. 

Let $\fbeta=(\beta_1,\beta_2,\ldots,\beta_m)$. We say that two vectors ${\bf j}(q)$ and ${\bf j}(q')$ produce the $\fbeta$-pair when for all $i$ 
if $j_i(q)=j_v(q')$ and $v\in[l_{i-1}+1;l_i]$, then $\beta_i=v$, otherwise $\beta_i=0$.

For example, if we have a function $\phi(x_1,x_2,x_3,x_4,x_5)$ and two samples $H_1$ and $H_2$, the first sample 
$H_1$ is for arguments $(x_1,x_2)$ (this means $l_1=2$) and the second sample $H_2$ is for arguments
$(x_3,x_4,x_5)$ (this means $l_2=5$). Now let we have resample vectors {\bf j}$(q)=(4,1,2,3,1)$ and {\bf j}$(q')=(1,2,2,4,3)$.
In this case, they will produce $\fbeta=(0,1,3,5,0)$-pair.

Note that the $\fomega$-pair is the specific case of the $\fbeta$-pair. $\fbeta$-pair becomes  $\fomega$-pair when only two 
variants are possible: $\beta_i=i$ or $\beta_i=0$. Also, zeros in the $\fomega$-pair are not stored.

The variance of obtained estimator can be calculated by formula (\ref{thetavar}). In order to calculate the covariance $Cov(\Theta^{*q},\Theta^{*q'})$ or the mixed moment $\mu_{11}$, we can use formula (\ref{cov}) or (\ref{mu11}), where 
we use the probability $P\{{\fbeta}\}$ of $\fbeta$-pair, conditional covariance $C({\fbeta})$ given $\fbeta$-pair and the conditional mixed moment $\mu_{11}({\fbeta})$ given $\fbeta$-pair.

Now let us show how to calculate the probability $P\{{\fbeta}\}$ of the $\fbeta$-pair. 
The arguments can be split into independent blocks. The block $i$ corresponds to arguments which belong to the sample $H_i$.
Let $\alpha_i$ be a number of non-zero elements at the block $i$. Then the probability $P(\fbeta)$ can be calculated using the
hypergeometrical distribution
\begin{equation}
\label{ex25_15}
P(\fbeta) = \prod_{i=1}^{k} \frac{\displaystyle{m_i\choose \alpha_i}{n_i-m_i\choose m_i-\alpha_i}}{\displaystyle{n_i\choose m_i}},
\end{equation}
Note that in the case when $n_i\le 2m_i$, it is impossible for two samples not to have common elements. For this case, let us put $\displaystyle{n\choose k}=0$, if $k>n$.

The calculation of $C(\fbeta)$ depends on the function $\phi$ structure. Note that
for the calculation of the probability $P(\fbeta)$ we need only the information about 
a number of non-equal elements in each block. In general, the information about the indices
of equal elements is used for a calculation of $C(\fbeta)$.

Now let us consider a specific case of the above described situation - the case when some elements are not simply equivalent, but their influence to the system work is equivalent \cite{Fio_00,Fio_02}. 
It can happen in the situation, when the function $\phi$ is commutative by the block arguments, i.e. changing the order of arguments inside the block does not change the function result. Note that a reliability function, which includes identical elements, can often be commutative by these arguments, because can include sum, min, max etc. of these arguments.

Here, we do not need to store for each element of ${\bf j}(q)$, what element of ${\bf j}(q')$ contains the same element of corresponding sample. We need only to know how much the same elements were selected for each block from the corresponding sample. So, instead of $m$-element $\fbeta$-pair we introduce the $k$-element $\falpha$-pair.

Formally, we say that two vectors ${\bf j}(q)$ and ${\bf j}(q')$ produce the $\falpha$-pair if for all $i$ 
$|\{j_{l_{i-1}+1}(q) \ldots j_{l_i}(q)\}\cap \{j_{l_{i-1}+1}(q') \ldots j_{l_i}(q')\}|=\alpha_i$, $i=1,\ldots,k$. This means that $\falpha$-pair stores only the number of common elements inside each block.
The probability $P(\falpha)$ can be calculated by formula (\ref{ex25_15}).

For example, if we have a function $\phi(x_1,x_2,x_3,x_4,x_5)$ and two samples $H_1$ and $H_2$, the first sample 
$H_1$ is for arguments $(x_1,x_2)$ (this means $l_1=2$) and the second sample $H_2$ is for arguments
$(x_3,x_4,x_5)$ (this means $l_2=5$). Now let we have resample vectors ${\bf j}(q)=(4,1,2,3,1)$ and ${\bf j}(q')=(1,2,2,4,3)$.
In this case, they will form $\falpha=(1,2)$-pair.



\subsection{The case of partially known distributions}

In the previous sections we supposed that distributions of all r.v. {\bf X}=\{$X_i,X_2,\ldots,X_m$\}
are unknown, but only sample populations {\bf H} are available for each variable. 
In many practical situations the distributions of some r.v. are known, 
but the distributions of the rest r.v. are unknown, and only sample
populations {\bf H} are available. The problem is, how to
use the available information about r.v. distribution in the most efficient way \cite{AndrF_99_3}.

Let $\phi$ be a known function of $m+\nu$ independent r.v.
{\bf X}=\{$X_1,X_2,\ldots,X_m$\} with unknown distributions, {\bf Z}=\{$Z_1,Z_2,\ldots,Z_\nu$\} with known distributions: $\phi({\bf X,Z})$.  The problem consists in
estimation of the expectation $\Theta=E\;\phi({\bf X,Z}).$

The idea of the estimation of $\Theta$ is the following: we use the conditional expectation of $\phi({\bf X,Z})$ given ${\bf X}$:
\begin{equation}
\label{condexp_partly}
g({\bf X})=E\;(\phi{\bf (X,Z)|X}).
\end{equation}

It is clear that $\Theta = E\;g({\bf X}).$

Two situations are possible for the conditional expectation ($\ref{condexp_partly})$: 
either $g({\bf X})$ has known functional form for any {\bf X} or 
the functional form of $g({\bf X})$ is unknown.

In the first case we can operate as in usual simple resampling, but we estimate $E\;g({\bf X})$ instead of
$E\;\phi({\bf X})$.
In each realization we create a resample $X^{*q}$ and use it to estimate $E\;g({\bf X})$:  
$\Theta^{*q}=g(X^{*q})$.
Then, the estimator $\Theta^*$ can be obtained by formula (\ref{thetaest}).

Now we consider the case when the conditional expectation (\ref{condexp_partly})
is unknown. In this case we are able to estimate $g({\bf X})$, 
because the distributions of random variables ${\bf Z}$ are
known. On the $q$-th step we generate $N$
mutually independent realizations $Z^{*q,i}$ of vector {\bf Z}, $i=1,\ldots,N$ and estimate $E\;g({\bf X})$ in the following way:
\begin{equation}
\label{rhoest_partly}
\Theta^{*q}=\frac{1}{N}\sum_{i=1}^{N}\phi(X^{*q},Z^{*q,i}).
\end{equation}

Then the estimator $\Theta^*$ can be obtained by formula (\ref{thetaest}).

Now let us see how hierarchical resampling can be applied in the case of
partially known distributions. Let the function $\phi$ be calculated by a calculation tree. The leaves of the tree correspond
to the variables $X_i$. Note that the intermediate functions $\phi_{m+1}, \phi_{m+2}, \ldots, \phi_{k}$
depend on the values of child nodes and the variables $Z_i$. 

Instead of the conditional expectation $g({\bf X})$, we must know the conditional distribution function $F_{v,X}(y)$ of each $\phi_v({\bf X,Z})$ given {\bf X}: $F_{v,X}(y)=P\{\phi_v({\bf X,Z}) \le y\}$. As in the case of the simple resampling, two variants are possible here: either the distribution function 
$F_{v,X}(y)$ can be calculated in each node $v$ or it is unknown.

Let us consider {\it the first variant} when during a sequential calculation of our
function we are able to find conditional distributions of its subfunctions.
Let $G_j(y)=P\{Z_j\le y\}$ is known distribution function
of $Z_j$. Then $F_{v,X}(y)$ can be calculated in
usual way:
\begin{equation}
\label{distrfun_partial}
F_{v,X}(y)=\mathop{\int\int\ldots\int}\limits_{\phi_v(X,Z)\le y} \prod_{j}dG_j(z_j).
\end{equation}

As the result, we have the following procedure. Let us consider the first (not initial) level.
We choose values from samples $H_i$, $i \in I^v$.
It allows us to calculate a realization $F^{q}_{v,X}(\cdot)$ of conditional distribution function $F_{v,X}(\cdot)$
by formula (\ref{distrfun_partial}). 
We repeat this procedure $n_v$ times and 
get a sample population $H_v$, which elements are the distribution
functions $F^{q}_{v,X}(\cdot)$, $q=1,2,\ldots,n_v$. 

Now let a vertex $v$ be an intermediate one. 
We have to extract the distribution functions $F_i$ instead
of simple values. Let $\{F_{i,X}^{*q}(\cdot)\}$, $i\in I^v$ be the set of the distribution
functions, extracted on the $q$-th step from the child samples.
Then we calculate the distribution function
\begin{equation}
\label{distrfun2}
F_{v,Y}^q(y)=\mathop{\int\int\ldots\int}\limits_{\phi_v(Y,Z)\le y}
\prod_{i} F_i^{*q}(y_i)\prod_{j}dG_j(z_j)
\end{equation}
and use it as $q$-th element of the sample $H_v$.

Finally for the root $k$ of the calculation
tree we calculate the estimator $\Theta^*$ by the following formula:
\begin{equation}
\label{thetaest3}
\Theta^* = \frac{1}{r} \sum_{q=1}^r\int\limits^{\infty}_0 y\,dF_{k,Y}^q(y),
\end{equation}
where $r=n_k$.

Now we consider {\it the second variant} when the conditional distributions
of the available subfunctions are unknown. 
For each vertex $v$ and
realization number $q=1,2,\ldots,n_v$, we have a sequence $X_{v,\xi}^q$,
$\xi=1,2,\ldots,N$ of $N$ independent realizations of subfunction
$\phi_v(\cdot)$ with the same distribution function $F_{v,Y}^q(\cdot)$.
Therefore, the $q$-th element of the sample population $H_v$ is the
vector ${\bf X}_{v}^q=\{X_{v,\xi}^q:\xi=1,2,\ldots,N\}$ that "represents"
unknown distribution $F_{v,Y}^q(\cdot)$.

On the $q$-th step we extract
a vector from each sample $H_i$, $i\in I_v$, forming resample $X_v^{*q}=\{X_i^{*q}\}$, $i\in I^v$. 
Then for each random variable $Z_i$ we generate (by the random
number generator) a vector of $N$ its independent realizations, forming vector of $Z$ realizations:   $Z_v^{*q}=\{Z_i^{*q}\};  Z_i^{*q}=\{Z_{i,\xi}^{*q}:\xi=1,2,\ldots,N\}$
in accordance to the distribution function $G_i(\cdot)$. Further we
calculate values $X_{v,\xi}^q=\phi_v(X_{v,\xi}^{*q},Z_{v,\xi}^{*q})$
and form a vector $X_{v}^q=(X_{v,\xi}^q:\xi=1,2,\ldots,N)$, which becomes $q$-th element of the sample $H_v$.

When the root $k$ of the calculation tree will be reached,
we are able to estimate $\Theta$ by analogy with (\ref{rhoest_partly}):
\begin{equation}
\label{form12}
\Theta^*=\frac{1}{rN}\sum_{q=1}^r\sum_{\xi=1}^N X_{k,\xi}^q,
\end{equation}
where $r=n_k$.

Note that previously $H_v$ has denoted as the sample
of function $\phi_v$ values. Now we have more general case: $H_v$ denotes
either the set of the conditional distributions $\{F_{v,Y}^q(\cdot)\}$ or the
set of vectors $\{X_{v}^q\}$ which represent these distributions.

Let us consider the same system as in the Section 1.2, but with
partially known distributions. Let us denote variables from that problem
as $X'_1, \dots, X'_6$. 
Let the distributions of random variables $X'_1$, $X'_3$ and $X'_5$ are unknown, 
but the distributions of $X'_2$, $X'_4$ and $X'_6$ are known.
In our present definitions, $X_1=X'_1$, $X_2=X'_3$, $X_3=X'_5$; $Z_1=X'_2$, $Z_2=X'_4$,
$Z_3=X'_6$. Our characteristic of interest 
$\Theta$ is the expectation of function
$$
\phi_t(x_1,x_2,x_3,z_1,z_2,z_3)=\left\{
\begin{array}{ll}
1 &\mbox{if}\;min\{max\{x_1,z_1\},min\{x_2,z_2\},x_3+z_3\}<t,\\
0 &\mbox{otherwise}.
\end{array}
\right.
$$

Here we have the case when 
the conditional expectation (\ref{condexp_partly})
is known.
It can be calculated easily:
$$
\label{ex3}
g_t(x_1,x_2,x_3)=\left\{
\begin{array}{ll}
1,&\mbox{if}\;x_2<t,\\
1-\overline{G_1}(t)\overline{G_2}(t)\overline{G_3}(t-x_3),&\mbox{if}\;x_2>t,x_1<t,x_3<t,\\
\overline{G_2}(t)\overline{G_3}(t-x_3),&\mbox{if}\;x_2>t,x_1>t,x_3<t,\\
\overline{G_1}(t)\overline{G_2}(t),&\mbox{if}\;x_2>t,x_1<t,x_3>t,\\
\overline{G_2}(t),&\mbox{if}\;x_2>t,x_1>t,x_3>t.
\end{array}
\right.
$$

Now we are able to use the resampling approach to estimate $\Theta^*$.
In this case, we have 8 $\fomega$-pairs, which are subsets of {\bf M}=\{1,2,3\}.
We can calculate the values $\mu_{11}(\fomega)$ and 
$P\{\fomega\}$.





\section{Resampling estimation of stochastic process parameters}

\subsection{The failure model with accumulation of damages}
\hyphenation{con-si-dered}
\hyphenation{dege-ne-ration}

\label{dsds}
Above we consider statistical models where time factor is absent. On the contrary
stochastic processes have a dynamic character. Here the efficiency investigation of resampling estimator encounters great difficulties \cite{Andr_00_2}. Let's consider some statistical models, which are implemented to different reliability problems.

The  model with accumulation of damages was considered in \cite{Afan_02} and \cite{Andr_06}.
It is based on the failure model and its modifications, presented in \cite{Andr_Gerc_72,Andr_94,Ger_00}. The model supposes two types of failures - initial and terminal failures. The initial failures (damages) appear according to a homogeneous Poisson process with rate $\lambda$. Each initial failure degenerates into a terminal failure after a random time $Z$. So if $i$-th initial failure appears at time $\tau_{i}$ then the corresponding terminal failure appears at the time instant $\tau_{i}+Z_{i}$. Terminal failure and the corresponding initial failure are eliminated instantly. We assume that $\{Z_{i}\}$ are i.i.d. r.v's, independent on $\tau_{i}$ with cdf $F(x)$. We take interest in the number of initial failures $X_{t}$ which did not degenerate into terminal failures at time $t$ (further initial failures) and the number of terminal failures $Y_t$ observed up to time $t$. Let $EX_{t}$ and $EY_{t}$ be the corresponding expectations, $P_{X_{t}}(i)=P\{X_{t}=i\}$, $P_{Y_{t}}(i)=P\{Y_{t}=i\}$ be the corresponding probability distributions, $i = 0, 1, \ldots $. 
\par
It is well known that $X_{t}$ and $Y_{t}$ are mutually independent r.v's, by that:
\begin{equation}
\label{ext}
EX_{t}=\lambda\int_{0}^{t}(1-F(x))dx,\  EY_{t}=\lambda\int_{0}^{t}F(x)dx,
\end{equation}

\begin{equation}
\label{ppxt}
P_{X_{t}}(i)=\frac{1}{i!}\left({EX_{t}}\right)^{i}exp(-EX_{t}), i=0,1, \ldots . 
\end{equation}

The probability $P_{Y_{t}}(i)$ is calculated analogously by formula ($\ref{ppxt}$) where $EX_{t}$  is replaced by $EY_{t}$. In future all formulas will be obtained for $EX_{t}$, but for  $EY_{t}$ all of them can be calculated analogously.

The rate $\lambda$ and the cdf $F(x)$ are unknown. Two samples are given: the sample $H_A$=($A_{1},A_{2},\ldots ,A_{n_{A}})$ of the intervals between initial failures and the sample $H_B$=($B_{1},B_{2},\ldots ,B_{n_{B}})$ of degeneration times. We need to estimate $EX_{t}$,$EY_{t}$,$P_{X_{t}}(i)$ and $P_{Y_{t}}(i)$ using samples $H_A$ and $H_B$.

In order to estimate values (\ref{ext}) and (\ref{ppxt}) we use the resampling approach. On the $q$-th realization we extract (without replacement) elements $A_{i_{1}(q)},A_{i_{2}(q)},\ldots $ from the sample $H_A$ obtaining the $q$-th resample  $A^{*q}=(A^{*q}_{1},A^{*q}_{2},\ldots,A^{*q}_{n_{A}} )$, where $A^{*q}_{k}=A_{i_{k}(q)}$. Then we calculate the instants of initial failures $\tau_1^{*q}=A_1^{*q}$, 
$\tau_2^{*q}=\tau_1^{*q}+A_2^{*q}$, $\ldots$\ . Then we extract $n_A$ values
$B_{j_{1}(q)},B_{j_{2}(q)},\ldots $ from the sample $H_B$ (suppose that $n_{A}\le n_{B}$), obtaining the $q$-th resample $B^{*q}=\{B^{*q}_{1},B^{*q}_{2},\ldots,B^{*q}_{n_{A}} \}$,
 where $B^{*q}_{k}=B_{j_{k}(q)}$.  This allows us to calculate the instants of terminal failures $\{
\tau^{*q}_{1}+B^{*q}_{1}, \tau^{*q}_{2}+B^{*q}_{2},\ldots,\tau^{*q}_{n_{A}}+B^{*q}_{n_{A}}\}$ 
and keep in mind the number of failures of each type up to time $t$. Further all extracted values are returned into the initial samples and the described procedure is reiterated $r$ times. 


Let $\zeta^{*q}_{j}$ be the indicator function of the event: "The $j$-th initial failure occurred, but did not degenerate into a terminal failure up to the time $t$ ":
\begin{equation}
\zeta^{*q}_j(t)=	\left \{ 
\begin{array}{ll} 
1 &\mbox{ if } \tau^{*q}_j \le t<\tau^{*q}_j+B^{*q}_j,\\
0 &\mbox{ otherwise}.
\end{array}
\right.
\end{equation}



Then the number of initial failures $X^{*q}_t$  at time $t$ for the $q$-th realization is $X^{*q}_t=\sum_{j=1}^{n_A}\zeta^{*q}_j(t).$

The resampling-estimator $E^*X_t$ of $EX_t$ can be obtained from formula ($\ref{thetaest}$), where $\Theta^{*q}=X^{*q}_t$.  


Now we need to calculate the resampling-estimators of the probabilities $P_{X_t}(i)$. Let $P^{*q}_{X_t}(i)$ be the indicator function of the event $\{X^{*q}_t =i \}$. The resampling-estimators $P^*_{X_t}(i)$ of the probabilities $P_{X_t}(i)$ can be determined by formula ($\ref{thetaest}$) taking $\Theta^{*q}=P^{*q}_{X_t}(i)$.

Let us calculate the expectations of the resampling-estimators (note that they are biased). Obviously $EP^{*}_{X_t}(i)=EP^{*q}_{X_t}(i)$, $EP^{*}_{Y_t}(i)=EP^{*q}_{Y_t}(i)$.  These expectations can be calculated taking into account the following reasoning: 1) The probability of the event that the number of initial failures occurred during time $t$ is equal to $j$ can be found by a Poisson distribution:	$	d_t(j)=\frac{(\lambda t)^j}{j!}exp(-\lambda t)$; 2)
	It is known, that if the number $j$ is fixed then the moments of initial failures are independent and uniformly distributed on (0, $t$); 3) If  $u \in (0,t)$ is the time moment of an initial failure appearance then in the time instant $t$ with probability $1-F(t-u)$ it remains initial; 4) The probability $p_1$ that at time $t$ the considered failure remains initial is:
$	p_1=\frac{1}{t}\int^t_0(1-F(t-u))du$.
	
	Therefore, the expectation $E(E^*X_t)$ of the resampling-estimator $E^*X_t$ is calculated as follows:
	
\begin{equation}
\label{eex}
	E(E^*X_t)=p_1\sum_{j=1}^{n_A}j d_t(j)+p_1 n_A
	\sum_{j=n_A+1}^{\infty}d_t(j).
\end{equation}

We also can find the expectation $EP^*_{X_t}$  of the estimator $P^*_{X_t}(i)$:

\begin{equation}
\label{epxt}
\begin{array}{c}
\displaystyle
	EP^*_{X_t}(i)=\sum_{j=i}^{n_A}d_t(j) {{j}\choose{i}}p^i_1
	(1-p_1)^{j-i}+ 
	\displaystyle
	{n_A \choose {i}}p_1^i(1-p_1)^{l-i}\sum_{n_A+1}^{\infty}d_t(j),\ i=1,2,\ldots .
	\end{array}
\end{equation}


Let us illustrate the idea of the calculation of the variance of $E^*X_t$.
It can be obtained from formula ($\ref{thetavar}$), where $\Theta^{*q}=X^{*q}_t$.  

For that purpose we need to calculate the covariance $Cov\left(X_t^{*q},X_t^{*q'}\right)$  for two different realizations $q$ and $q'$. Let us consider the second mixed moment $EX_t^{*q}X_t^{*q'}$. We have:
\begin{equation}
\label{extxt}
\begin{array}{c}
\displaystyle
EX_t^{*q}X_t^{*q'}=E\sum_{j=1}^{n_A}\zeta_j^{*q}(t)\sum_{i=1}^{n_A}\zeta_i^{*q'}(t)=
\sum_{j=1}^{n_A}\sum_{i=1}^{n_A}E\zeta_j^{*q}(t)\zeta_i^{*q'}(t).
\end{array}
\end{equation} 

Now we need to calculate $E\zeta_j^{*q}(t)\zeta_i^{*q'}(t)$. We have to take into account that $\zeta_j^{*q}(t)$  and $\zeta_i^{*q'}(t)$  can be formed by the same intervals between initial failures and by the same degeneration times. Let $\alpha_A$ be the number of the same intervals between initial failures on realizations $q$ and $q$'. Let $Z_B^j=i$  be the random event "the $j$-th initial failure in the $q$-th realization and the $i$-th initial failure in the $q$'-th realization have the same degeneration time." Then
\begin{equation}
\label{ezeta}
\begin{array}{ll}
\displaystyle
E\zeta_j^{*q}(t)\zeta_i^{*q'}(t)=&\displaystyle \sum_{\nu=0}^j P\{\alpha_A=\nu\}
\left(P\{R_B^j=i\}E(\zeta_j^{*q}(t)\zeta_i^{*q'}(t)|Z_B^j=i,\alpha_A=\nu)+\right.\\
&\left.(P\{R_B^j\ne i\})E(\zeta_j^{*q}(t)\zeta_i^{*q'}(t)|Z_B^j\ne i,\alpha_A=\nu)\right),\;j\le i.
\end{array}
\end{equation}

All elements of formula (\ref{ezeta}) can be easy calculated. It gives us a possibility to calculate the variance (\ref{thetavar}).

As an example, let us consider a Poisson flow of initial failures with rate  $\lambda$=0.5 and the triangular distribution of degeneration times with parameter $a$=2.
%

Table $\ref{tab1}$  presents the expectations $E\hat{P}_{X_t}(i)$  of the plug-in estimators $\hat{P}_{X_t}(i)$ and the expectations $EP^*_{X_t}$ of the resampling estimators  $P^*_{X_t}(i)$ (formula ($\ref{thetaest}$)) for the time $t$=5. The last column contains the real probabilities values $P_{X_t}(i)$  according to formula ($\ref{ppxt}$). The expectations are calculated for different numbers of initial failures $i$ and for different sample sizes $n_A$ (here $n_B=n_A$). We can see, that with increasing of $n_A$ the bias decreases, especially the bias of the resampling-estimator. 


\begin{table}[ht]
	\centering
		\caption{Expectations $E\hat{P}_{X_5}(i)$ of the Plug-in Estimators and $EP^*_{X_5}(i)$ of the Resampling-estimators}
		\label{tab1}
		\begin{tabular}{|l|l|l|l|l|l|l|l|}
			\hline
			$i$ & \multicolumn{2}{|c|}{$n_A=4$} & \multicolumn{2}{|c|}{$n_A=6$} & \multicolumn{2}{|c|}{$n_A=8$} & $ $\\
			\hline
			1&$E\hat{P}_{X_5}(i)$&$EP^*_{X_5}(i)$&$E\hat{P}_{X_5}(i)$&$EP^*_{X_5}(i)$&$E\hat{P}_{X_5}(i)$&$EP^*_{X_5}(i)$&$P_{X_5}(i)$\\
			\hline 
			0&.348&.370&.350&.368&.352&.368&.368\\
				\hline 
			1&.307&.374&.325&.368&.334&.368&.368\\
				\hline 
			2&.176&.189&.183&.184&.186&.184&.184\\
				\hline 
			3&.087&.058&.083&.062&.800&.061&.061\\
				\hline 
			4&.041&.009&.035&.015&.031&.015&.015\\
				\hline 
			5&.019& &.014&.003&.011&.003&.003\\
				\hline 
			6&.010& &.006& &.004& &.001\\
				\hline 
			7&.005& &.002& &.001& & \\
				\hline 	
		\end{tabular}
\end{table}

\begin{table}[ht]
	\centering
		\caption{Expectations, Variances and Mean Squared Errors of the Plug-in and Resampling Estimators of $E{X_5}$}
		\label{tab3}
		\begin{tabular}{|l|l|l|l|l|l|l|l|}
			\hline
			$i$ & $n_A=3$ & $n_A=4$&$n_A=5$ &$n_A=6$ & $n_A=7$ &$n_A=8$\\
			\hline
			$E(\hat{E}{X_5})$&1.41&1.32&1.25&1.21&1.19&1.16\\
				\hline 
				$E(E^*{X_5})$&0.89&0.96&0.99&0.997&0.99&0.99\\
				\hline 
			$Var\hat{E}{X_5}$&1.52&0.79&0.51&0.38&0.30&0.24\\
				\hline 
			$Var\ E^*{X_5}$&0.58&0.55&0.49&0.43&0.36&0.31\\
				\hline 
			$MSE\ \hat{E}X_5$&1.69&0.89&0.57&0.42&0.34&0.27\\
				\hline 
			$MSE\ E^*X_5$&0.59&0.55&0.49&0.43&0.36&0.31 \\
				\hline 
		\end{tabular}
\end{table}

Table $\ref{tab3}$ presents expectations  $E(\hat{E}X_5)$, $E(E^*X_5)$,variances $Var \hat{E}X_5$ ,$Var E^* X_5$ and mean squared errors  $MSE\  \hat{E}X_5$, $MSE\  E^* X_5$ for the estimators  $\hat{E}X_5$, $E^*X_5$. Note that the real values are $EX_5=1$. We can conclude that in many cases the resampling approach gives better estimators, if the criteria of efficiency are bias, variance or mean squared error. 

We can conclude, that the proposed resampling-approach is a good alternative to the traditional {\it plug-in} approach. Here the rate of convergence to the real values for the resampling-estimators is much more, than for the plug-in estimators. The only disadvantage of the resampling-approach consists in the impossibility to get good estimators of the probabilities $EP^*_{X_t}(i)$ for $i>n_A$. In this case it is better to use the plug-in estimators. Here it should to combine both approaches, using the resampling-estimators for $i< n_A$, the plug-in ones for $i> n_A$ and the normalization of the given probabilities. 

\subsection{The process of degradation - maintenance}
Those problems were considered by Afanasyeva(Fioshina) in \cite{Afan_05_1} and \cite{Afan_05_2}. 
Suppose we have two simple independent renewal processes {$X_i$, $i$=1,2,...} and {$Y_i$, $i$=1,2,...}, where {$X_i$} and {$Y_i$} are the sequences of nonnegative independent r.v., each with its common distribution function \cite{Cox_62,Ross_92}. Let $\displaystyle D_m=\sum_{l=1}^{m}{X_l}$ and $\displaystyle S_m=\sum_{l=1}^{m}{Y_l}$ be the times of the $m$-th renewal for corresponding processes. The cdf $F_X(x)$ and $F_Y(x)$ of sequences {$X_i$} and {$Y_i$} are unknown, but corresponding initial samples $H_X=\{X_1,X_2,\ldots,X_{n_X}\}$ and $H_Y=\{Y_1,Y_2,\ldots,Y_{n_Y}\}$ of sizes $n_X$ and $n_Y$ are available. Our purpose is the estimation of the probability $P\{D_m>S_k\}$, where $n_X\leq 2m$ and $n_Y\leq 2k$ . 

This problem has a lot of applications, for example, in reliability theory \cite{Lawless_02}. Let us consider the following degradation process.  The degradation level is increasing according to the degradation and decreasing according to the maintenance. If the degradation level becomes greater than the critical threshold $K$, where $K$ is a known integer, then the failure occurs. Our purpose is to estimate the failure absence probability for the $m$-th degradation moment. 

The described example can be considered in terms of renewal processes in the following way. Let the degradation corresponds to the first renewal process {$X_i$, $i$=1,2,...} and the time of the $m$-th renewal be the time of the $m$-th degradation.  Let the maintenance corresponds to the second renewal process {$Y_i$, $i$=1,2,...} and the time of the $m$-th renewal be the time of the $m$-th maintenance. Then the probability of interest, of the failure absence, is the probability, that the $m$-th degradation occurs later, that the $m-K$-th maintenance  $D_m>S_{m-K}$. It is also assumed, that the threshold level $K$ is known. 

Our task is to estimate the failure absence probability $P\{D_m > S_{m-K}\}$ that the $m$-th renewal of the degradation process {$X_i$} comes later, than the $m-K$-th renewal of the maintenance process {$Y_i$}.

In this case the function of interest is the indicator function $\phi$({\bf x,\bf y}) , where {\bf x}=$(x_1,x_2,\ldots,x_{m_X})$ and {\bf y}=$(y_1,y_2,\ldots,y_{m_Y})$ are vectors of real numbers: 
\begin{equation}
\label{lab1}
\phi(\bf{x,y}) =	\left \{ 
\begin{array}{ll} 
1 &\mbox{ if } \displaystyle\sum_{i=1}^{m_X}x_i > \displaystyle\sum_{i=1}^{m_Y}y_i,\\
0 &\mbox{ otherwise}.
\end{array}
\right.
\end{equation}


The resampling approach supposes the following steps. We choose randomly $m_X$ elements from the sample $H_X$ and $m_Y$ elements from the sample $H_Y$. The elements are taken without replacement, we remind that $n_X\geq 2m_X$, $n_Y\geq 2m_Y$. Then we calculate the corresponding value of function $\phi(\bf{x,y})$  using formula (\ref{lab1}). After that we return chosen elements into the corresponding samples.

We repeat this procedure during $r$ realizations. Let, like it was described in the first section ${\bf j}_{i}(q)=\{j^1_i(q),j^2_i(q),\ldots,j^{m_i}_i(q)\}$ be the indices of elements from the sample $H_i$, $i\in\{X,Y\}$ , that are chosen at the $q$-th realization.  Then for the $q$-th realization we obtain the following vectors: 
${\bf X}^{*q}=(X_{j^1_X(q)},X_{j^2_X(q)},\ldots,X_{j^{m_X}_X(q)}$), 
${\bf Y}^{*q}=(Y_{j^1_Y(q)},Y_{j^2_Y(q)},\ldots,Y_{j^{m_Y}_Y(q)}$).

The {\it resampling} estimator $\Theta^{*}$ can be obtained by formula ($\ref{thetaest}$) taking into account that the $q$-th resample estimator $\Theta^{*q}=\phi({\bf X}^{*q},{\bf Y}^{*q})$.  Obviously the estimator $\Theta^{*}$ is unbiased according to formula ($\ref{unbiased}$). We are interested in the variance of this estimator, witch can also be obtained by the formula ($\ref{thetavar}$), taking into account that $\Theta=E\;\phi({\bf X},{\bf Y})$. 

In order to estimate the variance of resampling-estimator, we have firstly to find the expression of the mixed moment $\mu_{11}=E\;\Theta^{*q}\Theta^{*q'}$ from the formula ($\ref{thetavar}$).

To calculate the moment $\mu_{11}$ the notation of $\falpha$-pairs can be used, which is the specific case of $\beta$-pairs, both of them were described in section 3.3. Here we suppose, that each resamples ${\bf X}^{*q}$ and ${\bf Y}^{*q}$ form their own blocks of sizes $m_X$ and $m_Y$ consequently. So, we have
$|\{j^1_i(q), \ldots ,j^{m_i}_i(q)\}\cap \{j^1_i(q'), \ldots ,j^{m_i}_i(q')\}|=\alpha_i$ for all $i\in \{X,Y\}$. It means, that for two different realizations $q$ and $q'$ resamples  ${\bf X}^{*q}$ and ${\bf X}^{*q'}$ have $\alpha_Y$ common elements and resamples ${\bf Y}^{*q}$ and ${\bf Y}^{*q'}$ have $\alpha_X$ common elements. The $\falpha$-pair consists on two elements $\falpha=(\alpha_X, \alpha_Y)$ and ${\bf j}(q)=\{{\bf j}^X(q), {\bf j}^Y(q)\}$. So to find $\mu_{11}$ by formula ($\ref{mu11}$) we have to calculate $P(\falpha)$ by formula ($\ref{ex25_15}$) and $\mu_{11}(\falpha)$. 
Now our task is to derive the formula for $\mu_{11}(\falpha)$ for this specific case. 
 
Let us include in sums  $D_{m_X}$ and $S_{m_Y}$ the upper index, that corresponds to the realization number: 
$D^q_{m_X}$ and  $S^q_{m_Y}$.
Then let's divide each sum into two parts, which consists of the common and the different elements of these sums for realizations $q$ and $q'$ (remind that they have $\alpha_X$ and $\alpha_Y$ common elements correspondingly):
\begin{equation}
\label{lab8}
\begin{array}{ll}
D_{m_X}^q=D_{m_X-\alpha_X}^{dif(qq')}+D_{\alpha_X}^{com(qq')},&
D_{m_X}^{q'}=D_{m_X-\alpha_X}^{dif(q'q)}+D_{\alpha_X}^{com(qq')},\\
S_{m_Y}^q=S_{m_Y-\alpha_Y}^{dif(qq')}+S_{\alpha_Y}^{com(qq')},&
S_{m_X}^{q'}=S_{m_Y-\alpha_Y}^{dif(q'q)}+S_{\alpha_X}^{com(qq')},\\
C_\alpha^{com(qq')}=D_{\alpha_X}^{com(qq')}-S_{\alpha_Y}^{com(qq')},&C_\alpha^{dif(qq')}=S_{m_Y-{\alpha_Y}}^{dif(qq')}-D_{m_X-{\alpha_X}}^{dif(qq')},\\
C_\alpha^{dif(q'q)}=S_{m_Y-{\alpha_Y}}^{dif(q'q)}-D_{m_X-{\alpha_X}}^{dif(q'q)}.
\end{array}
\end{equation}	

Therefore we can write: 
\begin{equation}
\label{lab9}
\begin{array}{c}
\mu_{11}(\falpha)=P\{\phi({\bf X}^{*q},{\bf Y}^{*q})=1,\phi({\bf X}^{*q'},{\bf Y}^{*q'})=1|\falpha\}=
\\
=P\{C_\alpha^{com(qq')}>C_\alpha^{dif(qq')},
C_\alpha^{com(qq')}>C_\alpha^{dif(q'q)}
\}=
\\
=\displaystyle\int^{+\infty}_{-\infty}P\{C_\alpha^{dif(qq')}<z,
C_\alpha^{dif(q'q)}<z\}dF_{com}(z|\falpha)=
\\
=\displaystyle\int^{+\infty}_{-\infty}F_{dif}(z|\falpha)^2 dF_{com}(z|\falpha),
\end{array}
\end{equation}
where $F_{com}(z|\falpha)$ is the cdf of $C_\alpha^{com(qq')}$, $F_{dif}(z|\falpha)$ is the cdf of $C_\alpha^{dif(qq')}$ and $C_\alpha^{dif(q'q)}$. They can be calculated by the following formula:
\begin{equation}
\label{lab10}
\begin{array}{c}
\displaystyle
  F_{com}(z|\falpha)= \displaystyle\int^{+\infty}_{-\infty}F_{X}^{(\alpha_X)}(x+z)dF_{Y}^{(\alpha_Y)}(x),\\
F_{dif}(z|\alpha)=
 \displaystyle\int^{+\infty}_{-\infty}F_{Y}^{(m_Y-\alpha_Y)}(x+z)dF_{X}^{(m_X-\falpha_X)}(x),
\end{array}
\end{equation}
where $F_{X}^{(m)}(x)$ is the cdf of r.v. $D_m$ and $F_{Y}^{(m)}(x)$ is a cdf of r.v. $S_{m}$.



Let's illustrate an example. Let r.v. \{$X_i$\} and \{$Y_i$\} have a normal distribution: $X_i\in N(\mu_X,\sigma_X)$, $Y_i\in N(\mu_Y,\sigma_Y)$. Then the sum $D_{m_X-\alpha_X}^{dif(qq')}$ from formula($\ref{lab8}$) 
has also normal distribution with the expectation $ED_{m_X-\alpha_X}^{dif(qq')}=(m_X-\alpha_X)\mu_X$ and the variance $VarD_{m_X-\alpha_X}^{dif(qq')}=(m_X-\alpha_X)\sigma_X^2$.
%
Analogously, the sum $S_{m_Y-\alpha_Y}^{dif(qq')}$ has also normal distribution with
expectation $(m_Y-\alpha_Y)\mu_Y$ and the variance $(m_Y-\alpha_Y)\sigma_Y^2$. Then cdf $F_{com}(z|\falpha)$ and $F_{dif}(z|\falpha)$ also have a normal distribution.

Let's  $X_i\in N(2,1)$, $Y_i\in N(2,1)$. Let our sample sizes be equal $n=n_X=n_Y$, and we consider the $m$-th degradation and different threshold levels $K=0 \ldots 3$. All calculations have performed for $r = 1000$ realizations.

We intend to compare the variance of estimators of resampling-approach with the mean squared error of classical approach. It is so because of resampling-approach estimators are unbiased, but classical ones on the contrary have bias.

\begin{table}[ht]
\centering
	\caption{Experimental results for Classical $\hat{\Theta}$ and Resampling $\Theta^*$ estimators}
	\label{tabb1}
		\begin{tabular}{|l|l|l|l|l|l|}
		\hline
		\multicolumn{2}{|c|}{}&$K=0$&$K=1$&$K=2$&$K=3$\\
			\hline
		$n=10,m=5$&$Var\;\hat{\Theta}$&.061&.043&.015&.002\\
			\cline{2-6}
			&$Bias\;\hat{\Theta}$&0&.028&.029&.013\\
			\cline{2-6}
			&$MSE\;\hat{\Theta}$&.061&.044&.015&.002\\
			\cline{2-6}
			&$Var\;\Theta^*$&.08&.055&.014&.001\\
			\hline
			$n=12,m=6$&$Var\hat{\Theta}$&.06&.045&.019&.004\\
			\cline{2-6}
			&$Bias\;\hat{\Theta}$&0&.028&.033&.019\\
			\cline{2-6}
			&$MSE\;\hat{\Theta}$&.06&.046&.02&.004\\
			\cline{2-6}
			&$Var\;\Theta^*$&.085&.058&.02&.002\\
			\hline
		\end{tabular}
\end{table}
In Table $\ref{tabb1}$ we can see the resampling-estimators' variance $Var\;\Theta^*$ comparing with classical approach estimators' variance $Var\;\hat{\Theta}$, bias $Bias\;\hat{\Theta}$, and mean squared error $MSE\;\hat{\Theta}$. The table shows how changes the results depending on different sample sizes $n$, degradation number $m$ and the threshold level $K$.

\section{Resampling Interval Estimation of Logical Systems}

In the previous sections we considered the point {\it resampling} estimators.
But in many applications we need to construct a confidence interval for
the system characteristics, not only a point estimator. A confidence
interval allows us better understand the obtained result and its accuracy.

Last  years the bootstrap approach has been applied 
for confidence interval construction 
\cite{David_97},  \cite{Dic_96}. 
In this section the {\it resampling} approach is used for this aim 
\cite{Andr_00,Andr_01,AndrF_04}. 
As an example, a confidence interval calculation for characteristics of 
logical system is considered. 

Let a function $\phi(X)$, does not depend on 
exact values of arguments, but on ordering  of these values only. 
This function can include boolean operators, 
comparisons, calculation of order statistics, including min and max, etc.
It can have two possible results only: one and zero.

Our aim is to construct an upper confidence interval $(a;\infty)$ for the 
expectation $\Theta = E\;\phi(X)$, that corresponds to
the confidence probability $\gamma$:
\begin{equation}
\label{confid}
P\{a < \Theta < 1\}=\gamma.
\end{equation}

Note that we are able to consider  our function $\phi(X)$ 
as function of permutations $\bpi\in\bPi$, where $\bPi$ 
is set of all permutations of elements $1, 2, \ldots, m$. 
In order to illustrate function dependence on permutation $\bpi$, 
we will write $\widetilde{\phi}(\bpi)$. We denote $\bPi_1$ 
a subset of permutations 
where $\widetilde{\phi}(\bpi)=1$, $\bPi_0$ a subset of permutations where 
$\widetilde{\phi}(\bpi)=0$: 
$\bPi_1 = \{\bpi\in\bPi: \widetilde{\phi}(\bpi) = 1\}$, 
$\bPi_0 = \{\bpi\in\bPi: \widetilde{\phi}(\bpi) = 0\}$. 
Therefore  our parameter of interest can be written as $\Theta=P\{\bpi\in\bPi_1\}$.

The procedure of the interval (\ref{confid}) construction is following. 
We estimate $\Theta$ using the resampling approach, obtaining estimates $\Theta^*$ by formula (\ref{thetaest}).
We repeat the resampling procedure $k$  times, obtaining the sequence of estimates
$\Theta^*_1,\Theta^*_2,\ldots,\Theta^*_k$. Then we order this sequence, obtaining the order statistics
$\Theta^*_{(1)},\Theta^*_{(2)},\ldots,\Theta^*_{(k)}$  and corresponding 
$\alpha$-quantile $\Theta^*_{(\lfloor\alpha k\rfloor)}$ of their distribution, 
where $\lfloor\alpha k\rfloor=\max\{\xi=1,2,\ldots : \xi\le\alpha k\}$. 
We set $\alpha=1-\gamma$  and the border of interval (\ref{confid}) becomes $
a=\Theta^*_{(\lfloor \alpha k\rfloor)}$,
so $(\Theta^*_{(\lfloor \alpha k\rfloor)},1)$ is accepted 
as $\gamma$-confidence upper interval for the true value of $\Theta$.

Due to the dependence between estimates $\Theta^*_1,\Theta^*_2,\ldots,\Theta^*_k$ the
true coverage probability of the constructed interval will differ from $\gamma$.
Our aim is to calculate the true value 
of covering probability $R$:
\begin{equation}
R=P\{\Theta^*_{(\lfloor\alpha k\rfloor)}\le\Theta\}. 
\end{equation}

How can we describe the total sample $H_1\cup H_2\cup \ldots\cup H_m$ 
after its ordering? Let $X_{(1)} \le X_{(2)} \le\ldots\le X_{(n)}$ be ordered 
sequence of  elements of $H_1\cup H_2\cup \ldots\cup H_m$. 
It is possible to  use  $n$-dimensional vector  ${\bf W} = (W_1, W_2, \ldots, W_n)$,
where $W_j\in\{1, 2, \ldots, m\}$ and $W_j = i$ means that  
element $X_{(j)}$ belongs to $H_i$. 

For example, let $m=3$, ${\bf X}_1 = \{2.5, 6.3, 1\}$, 
${\bf X}_2 = \{0.5, 2.1, 5.3,  5.2, 0.9\}$, 
${\bf X}_3 = \{6.1, 2.3\}$.
So, then $n = 10$, and ordered sequence is
$\{0.5, 0.9, 1, 2.1, 2.3, 2.5, 5.2, 5.3, 6.1, 6.3\}$,
$W = \{2, 2, 1, 2, 3, 1, 2, 2, 3, 1\}$.

Another way to describe this ordering is protocol notion 
introduced by Andronov in \cite{Andr_02}. The definition below generalizes the
protocol notion for our case.

Let 
$(x_1^{(1)},x_2^{(1)},\ldots,x_{n_1}^{(1)})$, 
$(x_1^{(2)},x_2^{(2)},\ldots,x_{n_2}^{(2)})$,\ldots,
$(x_1^{(m)},x_2^{(m)},\ldots,x_{n_m}^{(m)})$
are real-valued vectors and 
$\bx_1=(x_{(1)}^{(1)},x_{(2)}^{(1)},\ldots,x_{(n_1)}^{(1)})$, 
$\bx_2=(x_{(1)}^{(2)},x_{(2)}^{(2)},\ldots,x_{(n_2)}^{(2)})$,\ldots,
$\bx_m=(x_{(1)}^{(m)},x_{(2)}^{(m)},\ldots,x_{(n_m)}^{(m)})$
are corresponding ordered sequences: 
$x_{(1)}^{(i)}\le x_{(2)}^{(i)}\le\ldots\le x_{(n_i)}^{(i)}$, 
$i = 1, 2, \ldots, m$. We call $(n_2+1)$-dimensional vector 
${\bf C}(1) = (c_0(1), c_1(1), \ldots,  c_{n_2}(1))$,  
$c_0(1) + c_1(1) + \ldots + c_{n_2}(1) = n_1$ 
{\it a subprotocol of the first level}, 
where $c_j(1) = \#\{x_\nu^{(1)}\in \bx_1: x_{(j)}^{(2)} < x_\nu^{(1)}\le x_{(j+1)}^{(2)}\}$,
$j = 0, 1, \ldots, n_2$, $x_{(0)}^{(2)}=-\infty$,  $x_{(n_2+1)}^{(2)}=\infty$,
and $\#M$ means power of a set $M$. 
{\it A subprotocol of the $l$-th level} 
$\bC(l)$, $l = 2, 3, \ldots, m-1$,  is determined analogously 
using union $\bx_1\cup \bx_2\cup \ldots \cup \bx_l$ of sets 
$\bx_1, \bx_2, \ldots, \bx_l$ instead of $\bx_1$, and $\bx_{l+1}$ instead of $\bx_2$. 
We call  a sequence of subprotocols {\it a protocol} 
$\bC = (\bC(1), \bC(2), ... , \bC(m-1))$.

{\sloppy
For the previous example we have $\bC(1) = (0, 0, 1, 1, 0, 1)$, 
$\bC(2) = (4, 3, 1)$, $\bC = (\bC(1), \bC(2))$.
Obviously    protocol $\bC$ and vector ${\bf W}$   have one-to-one mapping. 
Often we prefer the protocols, because they can be recursively calculated. 

}

Now the unconditional coverage probability can be calculated 
as follows: $R=\sum_\bC P_\bC R_\bC$.

The probability $P_\bC$ to get a fixed protocol  $\bC$ can be constructed recurrently,
using recursion by the elements of the protocol $\bC$. If the protocol $\bC$ is fixed,
then the conditional probability $q_\bC$ of the event $\{\phi(X) = 1\}$ 
given the protocol $\bC$ is given by formula
\begin{equation}
q_\bC=\frac{\sum_{\bpi\in\bPi_1}h_\bpi(\bC)}{n_1 n_2\ldots n_m}.
\label{eq10}
\end{equation}
where $h_\bpi(\bC)$ is a number of  resamples 
that correspond to the permutation $\bpi$. 

The conditional probability of the event $\Theta^*_i<\Theta$ can be calculated by formula
\begin{equation}
\rho_\bC=P_\bC\{\Theta^*_i<\Theta\}=\sum_{\xi=0}^{\Theta r-1}\left(
\begin{array}{c}
r\\\xi
\end{array}
\right)
q_\bC^\xi(1-q_\bC)^{r-\xi}.
\label{eq18}
\end{equation}

Now we can calculate conditional probability to cover 
the true value of $\Theta$:
\begin{equation}
R_\bC=P_\bC\{\Theta^*_{(\lfloor\alpha k\rfloor)}\le\Theta\}=\sum_{\xi=\lfloor\alpha k\rfloor}^k
\left(
\begin{array}{c}
k\\\xi
\end{array}
\right)
\rho_\bC^\xi(1-\rho_\bC)^{k-\xi}.
\label{eq19}
\end{equation}

Let us consider an example. Let we have $m$-element sequential system. 
Our parameter of interest 
is the probability that the concrete element (for example, with index $m$) 
will fail first:
\begin{equation}
\label{ex61_1}
\Theta=P\{X_m=min(X_1,X_2,\ldots,X_m)\}.
\end{equation}

We need to construct the upper confidence interval for $\Theta$ with a given 
confidence level $\gamma$.

{\sloppy
\hbadness=2000
Let us use the {\it resampling} approach for this task. In this case the function $\phi(X)$ can be represented as
\begin{equation}
\label{ex61_2}
\phi(x_1,x_2,\ldots,x_m)=
\left\{
\begin{array}{ll}
1,& \mbox{if } x_m=min(x_1,x_2,\ldots,x_m),\\
0,& \mbox{otherwise}.
\end{array}
\right.
\end{equation}

}

Let $m=3$; let variables $X_1, X_2$ and $X_3$ have exponential distribution
with parameters $\lambda_1=3$, $\lambda_2=3$ and $\lambda_3=2$. 
In this case true value
of $\Theta=0.25$. Let also the number of experiments $k=10$ and 
the number of trials in each experiment $r=16$.
The results are presented in the table \ref{t6_1}.

\begin{table}[ht]
\caption{Actual coverage probability $R$ depending on sample sizes}
\label{t6_1}
\centering
\begin{tabular}{|c|c|c|c|c|c|}
\hline Sample sizes
&\multicolumn{5}{c|}{Requested coverage probability,}\\
$(n_1,n_2,n_3)$&$\gamma$=0.5	&$\gamma$=0.6    &$\gamma$=0.7	&$\gamma$=0.8	&$\gamma$=0.9\\
\hline
(3,3,3)		&0.533	&0.576	&0.625	&0.686	&0.770\\
\hline
(9,9,3)		&0.519	&0.571	&0.630	&0.701	&0.793\\
\hline
(4,4,4)		&0.521	&0.578	&0.640	&0.709	&0.797\\
\hline
(6,6,4)		&0.516	&0.576	&0.642	&0.715	&0.807\\
\hline
(5,5,5)		&0.515	&0.579	&0.646	&0.722	&0.817\\
\hline
(3,3,8)		&0.516	&0.581	&0.651	&0.728	&0.823\\
\hline
(4,4,7)		&0.512	&0.580	&0.652	&0.732	&0.830\\
\hline
\end{tabular}
\end{table}

\section*{Conclusion}

We considered various applications of resampling approach to reliability problems. This approach has a number of advantages as it allows to get unbiased estimators for characteristics of interest. It seems to us that resampling approach has a perspective future in the statistics and reliability.

\bibliography{biblio}
\end{document}